\documentclass[10pt,twocolumn]{article}
\usepackage[english]{babel}
\usepackage[activate={true,nocompatibility},final,tracking=true,kerning=true,spacing=,factor=1100,stretch=10,shrink=10]{microtype}
\usepackage{authblk} 
\usepackage[]{siunitx}
\usepackage{amsmath,bm}
\usepackage{graphicx}
\usepackage[nomove,nospace]{cite}
\usepackage{float}
\usepackage[all]{nowidow}
\usepackage[font=small]{caption}
\usepackage[left=48pt, right=42pt,top=46pt,bottom=60pt,headheight=15pt,headsep=10pt,letterpaper,twoside]{geometry}
\usepackage{titlesec}
\usepackage{lineno}
\usepackage{caption}
\usepackage{mathtools}
\usepackage[shortcuts]{extdash} 
\usepackage[dvipsnames]{xcolor}
\usepackage[colorlinks=true,linkcolor=black,citecolor=black,urlcolor=black]{hyperref} 
\author[1]{Johannes Bütow}
\author[1,2,3]{Jörg S. Eismann}
\author[1]{Varun Sharma}
\author[1]{Dorian Brandmüller}
\author[1,2,3,$^*$]{Peter Banzer}
\affil[1]{Institute of Physics, University of Graz, NAWI Graz, Universitätsplatz 5, 8010 Graz, Austria}
\affil[2]{Max Planck Institute for the Science of Light, Staudtstr. 2, 91058 Erlangen, Germany}
\affil[3]{Institute of Optics, Information and Photonics, University Erlangen-Nuremberg, Staudtstr. 7/B2, 91058 Erlangen, Germany}
\affil[*]{peter.banzer@uni-graz.at}
\date{} 
\DeclareCaptionLabelSeparator{bar}{\space\textbar\space}

\titleformat{\section}{\centering\normalfont\fontsize{12}{15}\bfseries}{\thesection}{0.5em}{}
\title{Generating free-space structured light with programmable integrated photonics} 
\makeatletter 
\renewcommand*{\@biblabel}[1]{\hfill#1.}
\makeatother

\begin{document}
\maketitle

\noindent
\textbf{Abstract:}
Structured light is a key component of many modern applications, ranging from superresolution microscopy to imaging, sensing, and quantum information processing.
As the utilization of these powerful tools continues to spread, the demand for technologies that enable the spatial manipulation of fundamental properties of light, such as amplitude, phase, and polarization grows further.
In this respect, technologies based on liquid\-/crystal cells, e.g., spatial light modulators, became very popular in the last decade. 
However, the rapidly advancing field of integrated photonics allows entirely new routes towards beam shaping that not only outperform liquid-crystal devices in terms of speed, but also have substantial potential with respect to robustness and conversion efficiencies.
In this study, we demonstrate how a programmable integrated photonic processor can generate and control higher-order free-space structured light beams at the click of a button. 
Our system offers lossless and reconfigurable control of the spatial distribution of light's amplitude and phase, with switching times in the microsecond domain.
The showcased on-chip generation of spatially tailored light enables an even more diverse set of methods, applications, and devices that utilize structured light by providing a pathway towards combining the strengths of programmable integrated photonics and free-space structured light.
\\

\section{Introduction}

Manipulating optical fields and locally shaping light's fundamental properties to meet specific needs has enabled breakthroughs on the fundamental research level as well as in advanced applications \cite{RubinszteinDunlop.2017,Forbes.2021,He.2022}.
Super-resolution microscopy \cite{Hell.2007,Maurer.2011}, communication \cite{Willner.2021}, optical-tweezers \cite{Volpe.2023}, metrology \cite{Neugebauer.2016}, and quantum information processing \cite{Mirhosseini.2015} are only a few amongst many important examples.
Numerous methods exist, each with its own advantages and disadvantages, that facilitate the generation of almost arbitrary optical fields and structured beams of light, as long as the generated fields are compliant with Maxwell's equations.
In many beam-shaping scenarios, amplitude, polarization or phase of a beam of light are sculpted by using liquid-crystal based devices \cite{Marrucci.2006,Zhang.2014,Lazarev.2019}.
Other techniques are based on metasurfaces, micromirrors, microelectromechanics or photonic crystals\cite{Avayu.2014,Shaltout.2019,Spagele.2021,Park.2021,Zhang.2022,Panuski.2022}.
Particularly, approaches based on integrated photonics have recently received great attention, owing to the fast-paced developments in this field.
Integrated photonic systems offer, for example, increased robustness and can readily incorporate other on-chip optical components like lossless splitters or laser sources \cite{Zhou.2023}, making all-integrated systems possible. 

Controlling, emitting and reconfiguring on-chip light precisely in real-time is also core to the emerging field of programmable integrated photonic circuits.
Here, meshes of universal 2$\times$2 optical gates (Mach-Zehnder interferometers) provide extensive and lossless control over the flow of light within microseconds \cite{Milanizadeh.2022}.
This enables applications in quantum information processing and the implementation of artificial neural networks \ \cite{Carolan.2015,Harris.2018,Taballione.2021} as well as matrix operations and communication \cite{Miller.2013c,Ribeiro.2016}.
Connecting an array of free-space emitters to these photonic circuits creates an interface to free-space light with control of relative amplitudes and phases \cite{Miller.2020}.
Such systems have enabled novel applications like on-chip separation and measurement of free-space modes \cite{Annoni.2017,Bogaerts.2020,Milanizadeh.2021,Milanizadeh.2022,Butow.2022}.
Emitting tailored on-chip light from free-space emitters is also the fundamental principle underlying optical phased arrays \cite{Sun.2013,Heck.2017}.
While receiving great attention in the past decade, applications mostly focused on beam steering and light detection and ranging systems \cite{Sun.2014,Yaacobi.2014,Abediasl.2015,Aflatouni.2015b,Rogers.2021} and less on free-space generation of special beams \cite{Sun.2014b,Notaros.2017,Guo.2021}.

Here we demonstrate the use of a programmable integrated photonic processor to generate free-space structured light on demand and in real-time.
Tailored on-chip fields, controlled losslessly within a mesh of universal 2$\times$2 optical gates, are fed into a square array of 16 grating couplers.
A desired free-space structured output is generated by calculating the settings of the calibrated photonic mesh appropriately. 
No further training is required.
We demonstrate, both numerically and experimentally the generation of various structured light beams and superpositions.
Despite a limited number of only 16 emitters, a wide variety of structured light fields are effectively generated, exhibiting exceptional modal quality.
This system paves the way towards a novel all-integrated platform for the generation of structured light, exhibiting excellent mode quality, extended functionality, no on-chip losses and high operating speeds.

\section{Main}

\begin{figure}[tb]
   \centering\includegraphics[width=1\columnwidth]{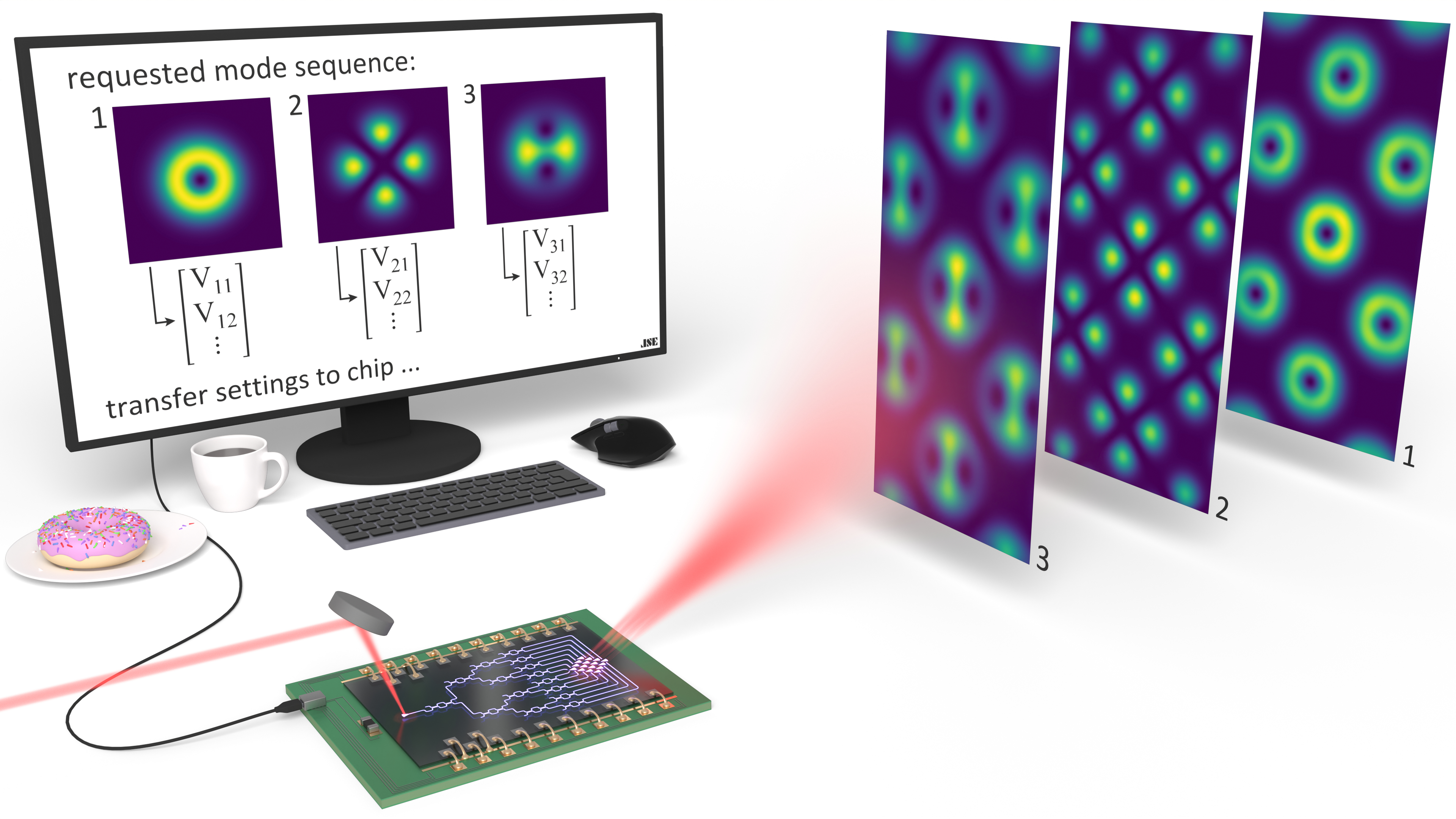}
   \captionsetup{name=Figure}
   \caption{Overview of the utilized system.
      A desired field distribution is specified on the computer, after which the required settings are calculated and sent to the photonic chip (bottom).
      These settings configure the photonic circuit such that it distributes the incoming coherent light across the mesh and subsequently re-emits it with tailored amplitude and phase, resulting in the desired field distribution in the far-field.
      Three exemplary desired structured modes are shown on the left, with the three respective theoretically generated free-space output fields on the right.}\label{fig:01figureArtisticSetup}
\end{figure}

We start by presenting an overview of the integrated photonics based system for the generation of structured light.
A schematic illustration is shown in Fig.\,\ref{fig:01figureArtisticSetup}.
For explanatory purposes, let us assume that the goal is to generate a sequence of three structured modes, desired for a certain application, shown in Fig.\,\ref{fig:01figureArtisticSetup} at the top left. 
After specifying these desired fields theoretically, the computer can calculate and store the exact control parameters that must be applied to the photonic chip in order to generate these structured modes.
The waveguide architecture on the photonic circuit, shown in the bottom center of Fig.\,\ref{fig:01figureArtisticSetup}, basically resembles a tree.
Coherent light of a laser source is coupled into a single waveguide at the beginning of the circuit, analogous to the trunk of a tree.
As the light flows through the waveguide mesh, at each junction the control parameters applied via the computer dictate how the light is distributed into the branches.
Eventually, this enables precise manipulation of the relative intensity and phase of the light in each waveguide at the output of the photonic circuit.
After processing, the light is coupled again to free space through an array of outcouplers, interferes, and propagates to the far-field, where the desired field distribution takes shape.
The resulting far-field intensity pattern obtained for the exemplary sequence of three structured modes is illustrated on the right in Fig.\,\ref{fig:01figureArtisticSetup}.

\begin{figure*}[thb]
   \centering\includegraphics[width=1\linewidth]{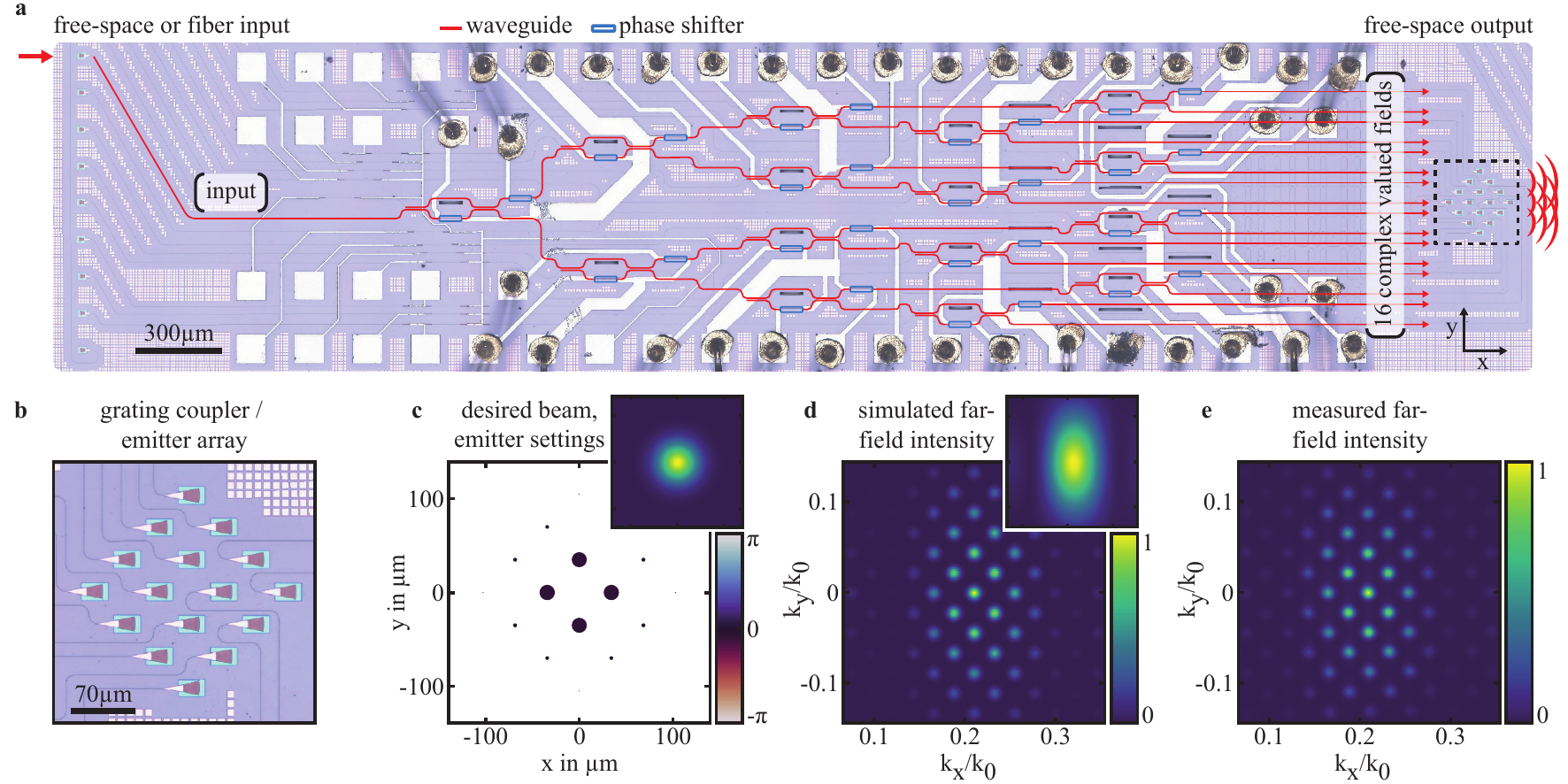}
   \captionsetup{name=Figure}
   \caption{Reconfigurable photonic integrated circuit generating structured light.
      \textbf{a},~Microscope image of the photonic chip, showing the waveguide mesh resembling a binary-tree of Mach-Zehnder interferometers.
      \textbf{b},~Detailed image of the 4$\times$4 square array of free-space emitters.
      \textbf{c}-\textbf{e},~Example of a generated array of Gaussian-like beams.
      \textbf{c},~Relative on-chip target intensities (indicated by size of circles) and phases (color-coded).
      The desired Gaussian intensity distribution to be generated is shown as inset at the top right.    
      \textbf{d},~Resulting theoretical far-field intensity distribution. Far-field intensity of a single emitter shown as inset.      
      \textbf{e},~Far-field intensities recorded experimentally with a camera.
      }\label{fig:02figureChip}
\end{figure*}

~\\
The underlying photonic architecture is based on a mesh of 15 reconfigurable Mach-Zehnder interferometers, arranged into a binary-tree, as shown in Fig.\,\ref{fig:02figureChip}a.
Each on-chip interferometer has two 50:50 beam splitters (\SI{3}{\decibel} directional couplers) and two heaters controlling the relative phase 
of the light propagating through the corresponding waveguide. 
By carefully adjusting these phases, each interferometer can split its single input field into two output fields of arbitrary relative intensity and relative phase.
After travelling through four stages of interferometers, the original input field coupled to the first waveguide is thus converted into 16 complex valued on-chip fields (see Fig.\,\ref{fig:02figureChip}a, right) of tailored relative intensities and phases.
It is worth noting that this field conversion does not introduce any fundamental loss, other than imperfections of the on-chip components \cite{Miller.2020}.
Subsequently, the generated fields are guided by their associated waveguides to a 4$\times$4 square array of emitters, i.e., standard grating couplers, acting as the free-space output interface that is shown in Fig.\,\ref{fig:02figureChip}a\,(right)\,and\,b.

An essential aspect in the usage of the photonic chip is determining the 16 complex valued on-chip field amplitudes that lead to the generation of a desired far-field distribution.
In this context, it is important to understand that the spatial structure of the fields emitted by an individual grating coupler is dictated by its design; thus it cannot be changed after manufacturing.
Tuning the applied phase shifts only affects the intensity and phase of each emitter, influencing its individual contribution to the far-field.
The near-field of a single grating coupler is obtained by means of a finite-difference time-domain simulation.
To quantify the degree to which this emitter field resembles the local field of the desired beam, it is necessary to calculate overlap integrals.
More precisely, the overlap integral between the field distribution of the target beam and the individual grating coupler output field is calculated in a plane just above the chip surface at a distance of \SI{300}{\nano\metre}.
Calculating the overlap integrals separately for each grating coupler results in 16 complex values.
They serve as the complex coefficients for the 16 on-chip fields that must be generated by the photonic circuit.
For the specific case of a Gaussian beam, the desired field distribution and the calculated values mentioned above are depicted in Fig.\,\ref{fig:02figureChip}c.
Note that in these calculations the target beams are always computed with a mean propagation direction being tilted along $x$-direction by \SI{12}{\degree} to the surface normal of the chip, since this is the approximate emission angle of the grating couplers. 
In all illustrations of the overlap values, though, the phase ramp associated with this tilt is subtracted in order to simplify the interpretation of patterns.

Now, we need to discuss how to control the photonic circuit to generate the desired mesh output.
Before this can be done, it is necessary to calibrate the chip carefully.
For this purpose, a calibration strategy reported recently\cite{Butow.2022} was adapted. 
In short, this calibration approach requires a known input beam, here a collimated Gaussian beam, sent onto the array of grating couplers used for beam shaping in the next steps (Fig. \ref{fig:02figureChip}a, right), which will couple into the circuit and propagate backwards through the waveguide mesh. 
During this process, the intensities emitted by all grating couplers on the other side (Fig. \ref{fig:02figureChip}a, left) are recorded.
Ultimately, analyzing the recorded data in this process allows for characterization of all relevant components in the circuit.
This comprises the splitting ratio of each on-chip beam splitter, coupling losses of the grating couplers, and the individual voltage-to-phaseshift relation of each thermal phase shifter. 
The latter is particularly important, since it allows for converting the applied voltages to resulting phase shifts, and vice versa, with heater-specific look-up tables.
The key advantage of using a calibrated chip is that from here onwards, the required settings of the chip to generate an arbitrary output distribution can be obtained by calculation.
The mathematical framework underlying the described process is explained in the methods section.

As one of the last steps, it must be understood how the fields emitted by a single emitter, and subsequently by the emitter array, behave upon propagation to the far-field.
To this end, using theoretical Fourier optics, the far-field of a single emitter can be calculated from the near-field simulation mentioned earlier.
The resulting far-field intensity distribution is displayed as an inset in Fig.\,\ref{fig:02figureChip}d, 
where the maximum light intensity is observed at an angle of approximately \SI{12}{\degree} $\left(\text{k}_\text{x}/\text{k}_\text{0}=0.21\right)$ relative to the chip's surface normal.
This distribution also reveals the available angular range where the system can generate structured light.
With the far\-/field of a single emitter at hand, calculating the total far\-/field of a coherent array of emitters is straightforward.
All individual far\-/fields are summed up, while using the complex amplitudes of the 16 on\-/chip output fields as coefficients and multiplying each contribution with a phase factor connected to the emitters position.
Utilizing the values depicted in Fig.\,\ref{fig:02figureChip}c, the resulting total far\-/field interference pattern is presented in Fig.\,\ref{fig:02figureChip}d, resembling an array of Gaussian-like beams.
The array-like properties of the generated pattern are determined by the spacing, number and individual angular emission spectra of the elementary emitters.
A corresponding measurement result is illustrated in Fig.\,\ref{fig:02figureChip}e, obtained by capturing the angular spectrum of the emitted light through a camera.

\begin{figure}[htb]
   \centering\includegraphics[width=1\columnwidth]{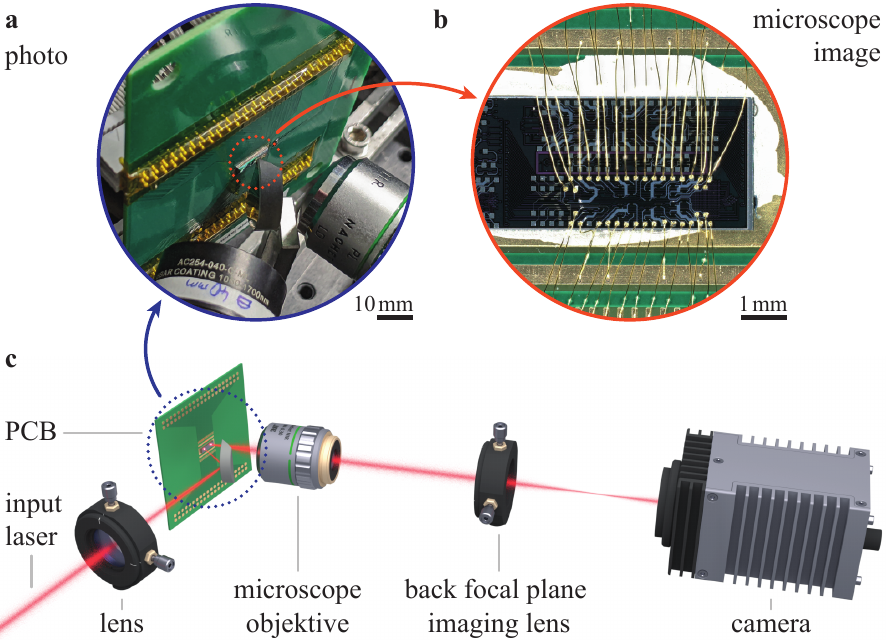}
   \captionsetup{name=Figure}
   \caption{Optical setup.
      \textbf{a}~Close-up photo of the printed circuit board (PCB) carrying the photonic chip that is highlighted by the dotted circle in the center. 
      \textbf{b}~Microscope image of the wire-bonded photonic chip on top of the PCB.
      \textbf{c}~Illustration of the experimental setup. 
      The incoming laser beam with a wavelength of \SI{1550}{\nano\meter} enters the image from the left and is free-space coupled into the photonic circuit via one grating coupler.      
      To measure the far\-/field of the emitted structured light, the back focal plane of the collection microscope objective is imaged onto a camera.
      }\label{fig:06setup}
\end{figure}

~\\
Before moving on to additional results, we briefly discuss the experimental optical setups used to capture the measured far-field data.
To first give an impression of the size, mounting, and wire-bonding of the photonic chip, a close-up photo of the PCB and a microscope image of the photonic chip are shown in Fig.\,\ref{fig:06setup}a-b, respectively.
An illustration of the optical setup that is build around the chip is shown in Fig.\,\ref{fig:06setup}c.
A free-space laser beam at a wavelength of \SI{1550}{\nano\meter} enters from the left and is focused by a lens via a D-shaped mirror onto one input grating coupler of the photonic chip.
Subsequently, the light that has passed through the photonic circuit and is re-emitted by the output grating-coupler array is collected and collimated by a microscope objective. 
The angular spectrum of the generated structured light is evaluated by imaging the back focal plane (Fourier plane) of the microscope objective onto a conventional infrared camera. 
One of the recorded camera images was already shown in Fig.\,\ref{fig:02figureChip}e, while additional images will follow.

\section{Results}

\begin{figure*}[!ht]
   \centering\includegraphics[width=1\linewidth]{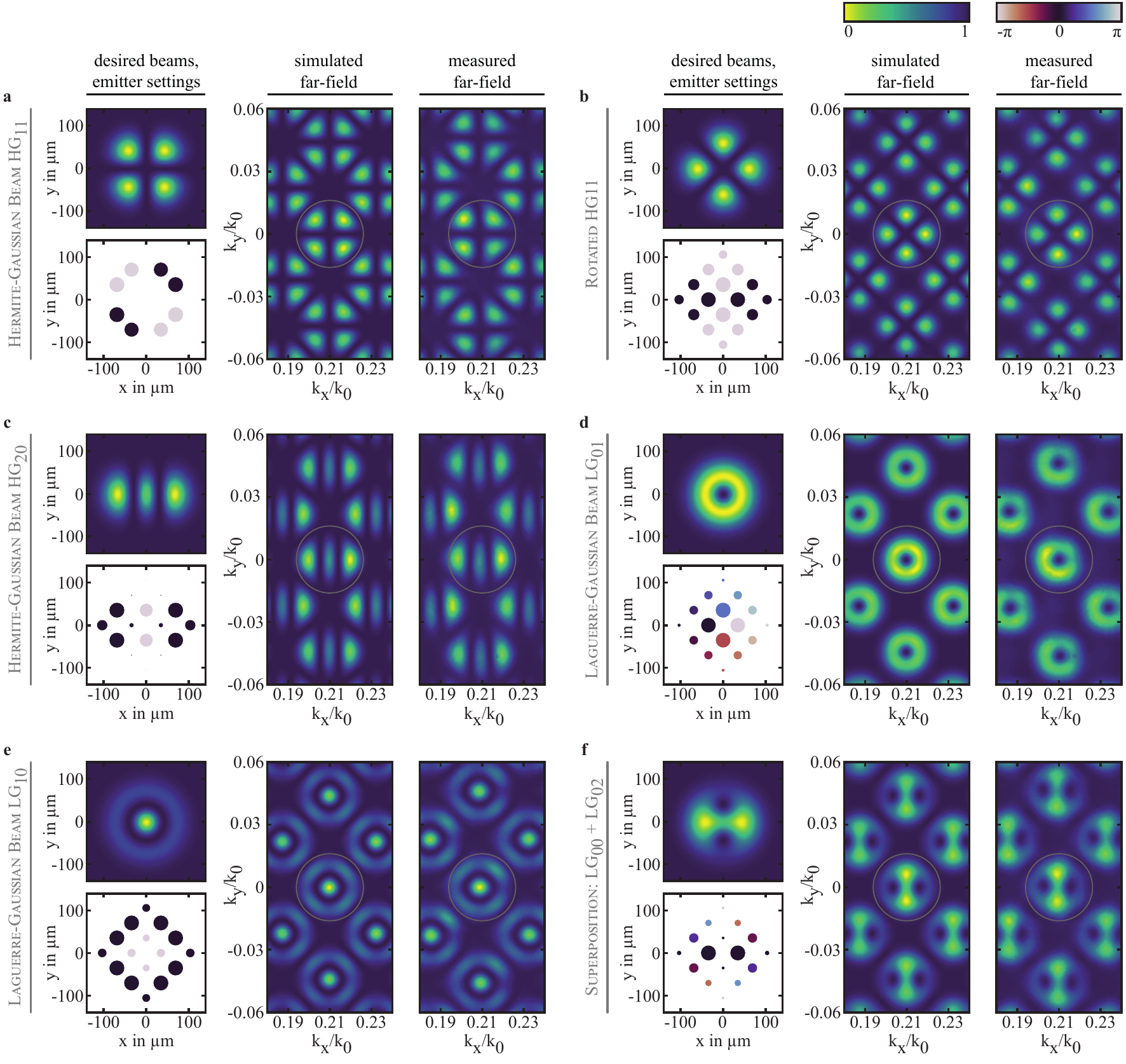}
   \captionsetup{name=Figure}
   \caption{Simulated and measured free-space far-field patterns of various structured beams.
      A target intensity distribution of light (top left) is sampled into 16 complex fields at the emitter positions.
      Individual intensities and phases are visualized as circles of corresponding size and color (bottom left).
      Coherently adding corresponding far-field emissions of the individual emitters results in the simulated array-like far-field pattern of the photonic chip (center).
      Only a region of the angular spectrum is shown here.
      A gray circle indicates the central diffraction order for reference.
      Experimentally, far-fields are recorded by imaging the angular spectrum of emitted light onto a camera (right).
      \textbf{a-b},~Arrays of Hermite-Gaussian beams of order $\text{HG}_\text{11}$ in different basis orientations.
      \textbf{c},~Additional example of Hermite-Gaussian beam generation.
      \textbf{d-e},~Arrays of Laguerre-Gaussian beams of different orders.
      \textbf{f},~Superposition of two Laguerre-Gaussian beams. More results are shown in the supplementary materials.}\label{fig:03figureResultsArtBoard1}
\end{figure*}

To showcase the system's far-reaching capabilities, we configured the photonic processor to produce a large variety of free-space patterns.
The main results comprising arrays of higher-order beams and modal superpositions are presented in Fig.\,\ref{fig:03figureResultsArtBoard1}.
The composition of all subfigures of Fig.\,\ref{fig:03figureResultsArtBoard1} is identical.
The desired intensity distribution is shown at the top left, with the corresponding intensity and phase settings of the emitters, i.e., the calculated overlap values, plotted below.
The simulated free-space far-field intensity distribution generated using this configuration is shown in the center, with the experimental counterpart depicted on the right.
The angular region containing the central diffraction order is encircled in gray.
The target distribution to be generated can be clearly identified within this area, with neighboring diffraction orders distributed as copies around it.

In Fig.\,\ref{fig:03figureResultsArtBoard1}a-b, we present the generation of an array of Hermite-Gaussian beams \cite{Svelto.2010} of order $\text{HG}_\text{11}$ in two different basis orientations.
It can be seen that in both cases the generated structured fields resemble the desired beams very well.
Upon close inspection, small differences in the shape of the outer parts of the intensity lobes can be identified between the desired beam and the generated far-fields in Fig.\,\ref{fig:03figureResultsArtBoard1}a.
Here, the features of the desired beam are not optimally sampled with the emitter array due to an unfavorable orientation of the modal basis, causing half of the available emitters to not contribute to the generation of the far-fields, a direct consequence of the small number of emitters in this study.
Generally, how well an arbitrary desired field distribution can be generated with the utilized system mainly depends on the adequate sampling of its features with the given array of emitters.
In the present example the generation of  beams of order $\text{HG}_\text{11}$ can be improved by rotating the modal basis.
In Fig.\,\ref{fig:03figureResultsArtBoard1}b a $\SI{45}{\degree}$-rotation results in a better match between the beams symmetry and the layout of the emitters.
This further increases the resemblance between the intensity patterns of the desired beam and the generated far-fields.
Fig.\,\ref{fig:03figureResultsArtBoard1}c shows the generation of an array of Hermite-Gaussian beams of order $\text{HG}_\text{20}$.
In Fig.\,\ref{fig:03figureResultsArtBoard1}d an array of Laguerre-Gaussian beams\cite{Allen.1992} $\text{LG}_\text{01}$ is generated.
Minor deviations can be observed in the measured far-fields, which might arise from, e.g., the calibration of the photonic mesh or differences in the emitters individual emission properties due to manufacturing.
Fig.\,\ref{fig:03figureResultsArtBoard1}e shows an array of Laguerre-Gaussian beams with radial index of\,1 $(\text{LG}_\text{10})$ being generated.
In Fig.\,\ref{fig:03figureResultsArtBoard1}f, we demonstrate a superposition of beams, i.e., a combination of a fundamental Gaussian ($\text{LG}_\text{00}$) and a Laguerre-Gaussian ($\text{LG}_\text{02}$) beam.
Equal relative amplitudes and no relative phase between the beams are chosen in this case.
Note that the generated far-field distributions are locally rotated with respect to the target distribution.
The physical origin of this rotation is the different Gouy phases acquired by different mode orders in the superposition as they propagate from the chip to the far field.
The resulting rotation depends on the individual mode contributions and could be accounted for in the experiments.

More examples of generated far-field distributions are shown in the Supplementary~Fig.\,\ref{fig:05figureResultsArtBoard3}-\ref{fig:04figureResultsArtBoard2} (resizing, refocusing, translation, additional higher-order modes and superpositions).
In addition, we show all these aspects in a Supplementary Video, where the chip's output is modulated, simulated, and recorded.
We also show a video in which we modulate and record the chip's output at high speeds, limited only by the frame-grabber of our camera (\SI{2}{\kilo\hertz}), but not by the photonic device itself.
With the utilized photonic chips, switching speeds down to microseconds would be possible.

\section{Outlook and Conclusions}

We have demonstrated and experimentally verified the generation and control of higher-order free-space structured light fields using a programmable integrated photonic processor.
Precise and lossless routing of light on the chip allows tailoring the relative amplitudes and phases of an emitter array with great flexibility, versatility and at exceptional high speeds.
Even with only a 4$\times$4 array of emitters, a wide variety of spatial higher-order modes and superposed beams can be generated with high quality.
This further extends the free-space applications of reconfigurable photonic integrated circuits and provides a powerful tool for existing applications involving structured light.

In the end, we would like to discuss some possibilities and prospects for future implementations of programmable integrated photonic processors used to generate structured light.
The overall number of emitters can be expected to increase soon.
Consequently, adequate sampling and subsequent generation of more complex distributions of light will become possible.
Meanwhile, the portfolio of novel integrated optical components and building blocks ready to be combined with photonic circuits is increasing continually.
Some developments, all directly applicable to our method of generating structured light, are particularly noteworthy.
For example, including polarization\-/sensitive grating couplers \cite{Su.2018, Nambiar.2018} into the emitter array would also allow future devices to structure the spatial polarization distribution.
Regarding the wavelength of operation, recent advances of photonic processors in silicon nitride \cite{Munoz.2019, PerezLopez.2021} already make a system like the one presented here possible in the visible spectral range.
The system discussed in this article operates in the infrared at \SI{1550}{\nano\meter}, the design wavelength of the given architecture, which could readily be changed by re-designing on-chip components appropriately.
In regard to the switching speed of the generated output fields, dynamic control within microseconds has been achieved, going beyond the capabilities of many established beam-shaping techniques.
This aspect could be improved even further by implementing alternative on-chip phase shifter technologies \cite{Aflatouni.2015b}, enabling the generation of structured light in space and time with bandwidths in the MHz domain.
Finally, a fully integrated system could be realized by implementing one or multiple on-chip controllable light sources \cite{Zhou.2023}.
This would result in a robust, portable, all-integrated structured light generator ready to be deployed in all sorts of structured light based applications.

\bibliographystyle{naturemag}
\footnotesize
\bibliography{modeGenerationBibliography}

\normalsize
\vspace{1cm}
\section*{Methods}

\noindent
\textbf{Integrated circuit design and fabrication}\newline
The photonic processor is based on a \SI{220}{\nano\meter} silicon-on-insulator platform and was fabricated commercially. 
All on-chip elements are standard foundry elements designed for operation at a wavelength of \SI{1550}{\nano\meter}.
The waveguides are \SI{500}{\nano\meter} wide. 
The on-chip 50:50 beam splitters (\SI{3}{\decibel} directional couplers) within the 2$\times$2 optical gates are \SI{40}{\micro\meter} long and feature a waveguide spacing of \SI{300}{\nano\meter}.
Two phase\-/shifters, i.e., thermal tuners of embedded TiN strips, actuate the reconfigurable Mach-Zehnder interferometers.
Driving voltages between \SIrange{0.2}{4}{\volt} enable a full $2\pi$ relative phase shift.
The grating couplers were initially designed for fiber-coupling to a transverse-electric (TE) polarized waveguide mode.
\vspace{.5cm}

\noindent
\textbf{Mesh configuration to generate arbitrary on-chip fields}\newline
To generate a desired free-space field distribution, the photonic chip has to be configured to create the associated 16 complex\-/valued on-chip fields leading into the emitter array.
Here we explain how to obtain the required settings for such a configuration of the interferometric mesh.
The desired 16 relative amplitudes and phases on the right serve as the initial parameters (compare Fig.\,\ref{fig:02figureChip}).
From there, the processing in each Mach-Zehnder interferometer can be calculated individually, progressing towards the left, from where the on-chip routing of light emanates from the single waveguide.

\begin{figure}[htb]
   \centering\includegraphics[width=1\columnwidth]{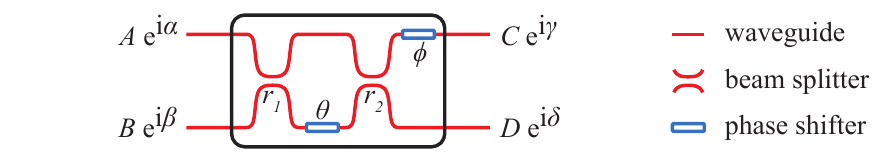}
   \captionsetup{name=Figure}
   \caption{Mach-Zehnder interferometer (2$\times$2 optical gate).
      Tuning the phase shifts $\theta$ and $\phi$ controls how light is processed upon transmission. 
      }\label{fig:07figureSingleMZI}
\end{figure}

In the following, we focus on a single Mach-Zehnder interferometer, illustrated in Fig.\,\ref{fig:07figureSingleMZI}, and demonstrate how to obtain the required phase shifter settings $\theta$ and $\phi$ to split an incoming field into two output fields of specific relative amplitude and phase.
As can be seen, each interferometer has two inputs, two outputs, two beam splitters, and two phase\-/shifters.
Input and output fields are described using a positive real\-/valued amplitude ($A, B, C, D$) and a phase term ($\alpha, \beta, \gamma, \delta$).
Moreover, each on-chip beam splitter has a field reflectivity $r$ and transmissivity $t = \mathrm{i}\sqrt{1-r^2}$.
For the purpose of generating structured light that is demonstrated here, light flows from left to right. 
In addition, due to the binary-tree waveguide architecture, there is always light only in one input waveguide, i.e., either A or B equals zero.

As an example, we consider an optical gate with light only in the lower input; thus, $A=0$.
This could correspond to any of the four right-most interferometers in the binary tree in Fig.\,\ref{fig:02figureChip}a.
As explained in the main part of the manuscript, the required output fields are dictated by the target field distribution. 
Accordingly, $C, D, \gamma, \delta$ are known, and the goal is to calculate $B, \beta, \theta, \phi$.
By following the paths light can take through the interferometer, the upper/lower output can be calculated as follows \cite{Miller.2020}:
\begin{align}
   C\:\mathrm{e}^{\mathrm{i} \gamma} &= B\:\mathrm{e}^{\mathrm{i} \beta}\left(t_1r_2+r_1t_2\:\mathrm{e}^{\mathrm{i}\theta}\right)\:\mathrm{e}^{\mathrm{i} \phi}, \label{eq:upperOutput} \\
   D\:\mathrm{e}^{\mathrm{i} \delta} &= B\:\mathrm{e}^{\mathrm{i} \beta}\left(t_1t_2+r_1r_2\:\mathrm{e}^{\mathrm{i}\theta}\right). \label{eq:lowerOutput}
\end{align}
Since there is no fundamental loss involved in this conversion, the input amplitude can be calculated through energy conservation
\begin{equation}
   B = \sqrt{C^2+D^2}.
\end{equation} 
Next, both sides of Eq.\,\eqref{eq:lowerOutput} are multiplied with their complex conjugates and solved for $\theta$
\begin{equation}
   \theta = \pm\:\text{acos} \left(\frac{ \left(\tfrac{D}{B}\right)^2 - r_1^2r_2^2 -t_1^2t_2^2}{2 r_1r_2t_1t_2}\right) \eqqcolon \pm\:\text{acos}\left(z_1\right). \label{eq:ampRatio}
\end{equation}
Both the positive and the negative sign will yield a true solution of $\theta$ to this problem. 
Without loss of generality, the positive one is selected, giving $\theta$ in the range between $0$ and $\pi$.
With $\theta$ now being fixed, Eq.\,\eqref{eq:lowerOutput} is rearranged to
\begin{equation}
   \mathrm{e}^{\mathrm{i} \beta} = \frac{ D\:\mathrm{e}^{\mathrm{i} \delta}}{ B (t_1t_2+r_1r_2\:\mathrm{e}^{\mathrm{i}\theta}) } \eqqcolon z_2.     
\end{equation}
This equation can be solved for the phase $\beta$ using the four-quadrant inverse tangent
\begin{equation}
      \beta = \text{atan2}\left(\operatorname{Im}\left(z_2\right),\operatorname{Re}\left(z_2\right)\right).
\end{equation}
Finally, with now also $\beta$ being fixed, the last two steps can be repeated to determine $\phi$ by rearranging Eq.\,\eqref{eq:upperOutput}
\begin{equation}    
   \mathrm{e}^{\mathrm{i} \phi} = \frac{ C\:\mathrm{e}^{\mathrm{i} \gamma}}{ B\:\mathrm{e}^{\mathrm{i} \beta}\:(t_1r_2+r_1t_2\:\mathrm{e}^{\mathrm{i}\theta}) } \eqqcolon z_3,     
\end{equation}
and solving for $\phi$
\begin{equation}
   \phi = \text{atan2}\left(\operatorname{Im}\left(z_3\right),\operatorname{Re}\left(z_3\right)\right).
\end{equation}
Reaching this point, the settings of the interferometer ($\theta, \phi$) and the required input field ($B\mathrm{e}^{\mathrm{i}\beta}$) are determined.
The calculation for the case that light enters the upper waveguide, and thus, ($B = 0$), is done analogous.
Finally, to calculate all 16 relative target amplitudes and phases of interconnected interferometers, each column of interferometers in the binary-tree mesh is solved progressively. 

A limit in achieving arbitrary output amplitudes comes from imperfect on-chip beam splitters.
Only interferometers with perfect 50:50 splitters (\SI{3}{\decibel} directional couplers), i.e., $|r|^2 = 0.5$, enable generating an arbitrary ratio of output amplitudes.
An interferometer with real world splitters can achieve a desired output amplitude ratio only if $|z_1| \leq 1$, compare Eq.\,\eqref{eq:ampRatio}.
Otherwise, the target amplitude ratio is too extreme and cannot be reached.
To still generate the ratio as good as possible we set $|z_1| = 1$ in this case.
For our experiments this limitation is negligible, since the reflectivity of all directional coupler is very close to the ideal value with $|r|^2 = \num{0.494\pm0.005}$.
The above calculations also reveal the required phase\-/shifting ranges to generate arbitrary output fields.
$\theta$ has to cover a range of at least $\pi$, either $\theta \in [0,\pi]$ or $\theta \in [-\pi,0]$.
$\phi$ has to cover the full range of $\phi \in [-\pi,\pi]$.
\vspace{.5cm}

\noindent
\textbf{Supplementary Video}\newline
Video of structured light being generated and dynamically controlled can be found at: \newline 
\url{https://cloud.uni-graz.at/s/nZZnZxymBBotexY}\newline
Video of high-speed generation of structured light can be found at: \newline
\url{https://cloud.uni-graz.at/s/bWLWwJGp5s8NW8T} \vspace{.5cm}

\footnotesize
\noindent
\textbf{Data availability}
The data that support the findings of this study are available from the corresponding author upon reasonable request.

\noindent
\textbf{Acknowledgements}
This work was supported by the European Commission through the Horizon 2020 Programme project (SuperPixels, 829116).
We thank all members of the SuperPixels consortium for fruitful discussions and collaboration.
We thank Maziyar Milanizadeh, Francesco Morichetti, Charalambos Klitis and Marc Sorel for the photonic circuit design, and Daniel Eisenkölbl for wire-bonding the photonic chip.\newline

\setcounter{figure}{0}
\begin{figure*}[!t]
   \centering\includegraphics[width=1\linewidth]{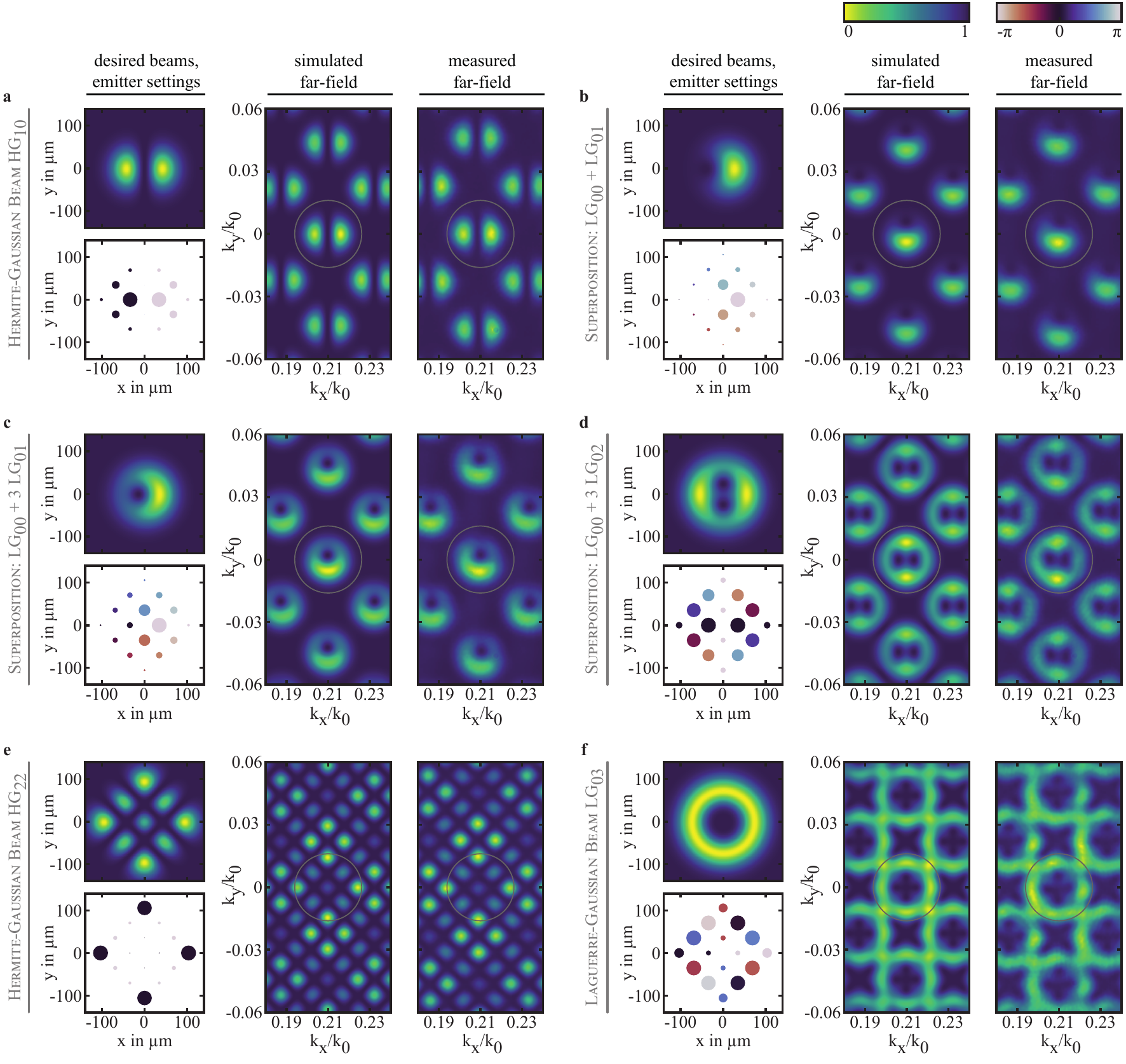}
   \captionsetup{name=Supplementary Figure}
   \caption{Additional results on generated structured light.
      Higher order beams and superpositions. 
      \textbf{a},~Generation of an array of Hermite-Gaussian $(\text{HG}_\text{10})$ beams.
      \textbf{b},~Superposition of Laguerre-Gaussian beams $\text{LG}_\text{00}$ and $\text{LG}_\text{01}$ of equal amplitudes.
      \textbf{c},~Superposition of Laguerre-Gaussian beams $\text{LG}_\text{00}$ and $\text{LG}_\text{01}$ of different amplitudes.
      \textbf{d},~Superposition of Laguerre-Gaussian beams $\text{LG}_\text{00}$ and $\text{LG}_\text{02}$ of different amplitudes.
      \textbf{e-f} Examples demonstrating the limits of the generation of structured light with only 16 emitters.
      \textbf{e},~Generation of an array of rotated Hermite-Gaussian $\text{HG}_\text{22}$ beams. 
      While most features of this target distribution still resemble well in the far field, it can be seen that individual diffraction orders come very close. 
      \textbf{f},~Generation of an array of Laguerre-Gaussian beams $\text{LG}_\text{03}$.
      The target distribution is not sufficiently sampled, resulting in overlapping diffraction orders. 
      Notably, this behavior is well predictable by theory, as simulation and experiment of the generated light are in excellent agreement.}\label{fig:05figureResultsArtBoard3}
\end{figure*}
\begin{figure*}[t]
   \centering\includegraphics[width=1\linewidth]{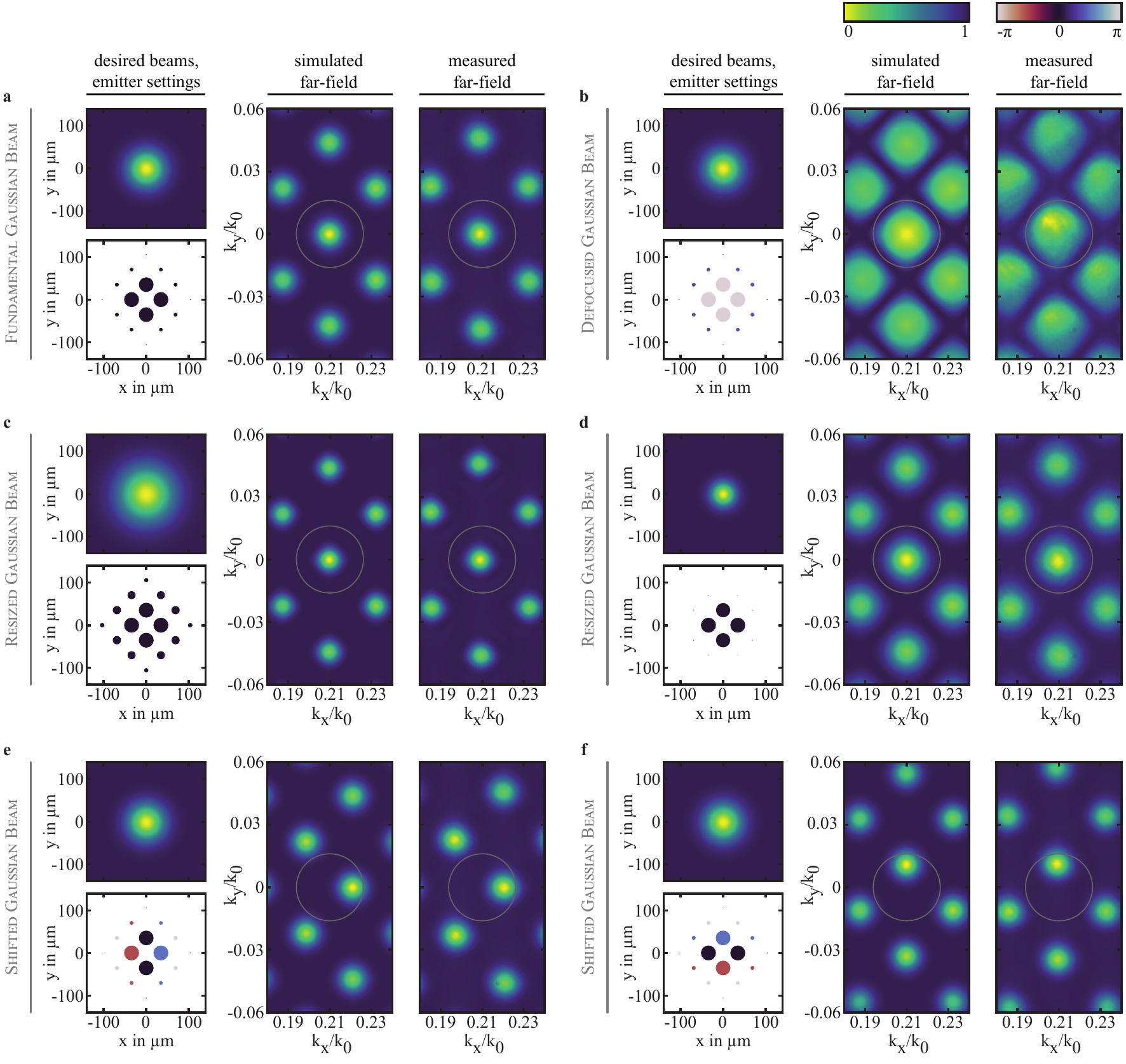}
   \captionsetup{name=Supplementary Figure}
   \caption{Additional results on generated structured light.
      Various modifications of a generated array of Gaussian beams.
      \textbf{a},~Detailed view of Gaussian beams from Fig.\,\ref{fig:02figureChip}c-e in the main text.
      \textbf{b},~Defocused Gaussian beams. An additional spherical phase is applied to the array of emitters.
      \textbf{c-d},~Resized Gaussian beams. Comparing these examples shows clearly that near-field and far-field beam widths are inversely related to each other.
      \textbf{e-f},~Dynamically shifted Gaussian beams. Additional linear phase ramps are applied to the array of emitters.}\label{fig:04figureResultsArtBoard2}
\end{figure*}
\end{document}